\documentclass[a4paper]{jpconf}
\usepackage{graphicx}
\usepackage{siunitx}

\begin{document}
\title{Status and Prospects for the AWAKE Experiment}

\author{Marlene Turner for the AWAKE Collaboration}

\address{CERN, Geneva, Switzerland }

\ead{marlene.turner@cern.ch}

\begin{abstract}
The AWAKE Collaboration is pursuing a demonstration of proton-driven plasma wakefield acceleration of electrons. The AWAKE experiment uses a \SI{400}{GeV/c} proton bunch from the CERN SPS, with a rms bunch length of $6$-\SI{15}{cm}, to drive wakefields in a \SI{10}{m} long rubidium plasma with an electron density of $10^{14}-10^{15}$cm$^{-3}$. Since the drive bunch length is much longer than the plasma wavelength ($\lambda_{pe}<$\SI{3}{mm}) for these plasma densities, AWAKE performed experiments to prove that the long proton bunch self-modulates in the plasma (2017). The next step is to demonstrate acceleration of electrons in the wakefields driven by the self-modulated bunch (2018). We summarize the concept of the self-modulation measurements and describe the plans and challenges for the electron acceleration experiments.
\end{abstract}

\section{Introduction}
The goal of AWAKE\cite{AWAKE,AWAKE2,AWAKE3}, the Advanced proton driven plasma WAKefield Experiment, is to show that plasma wakefields driven by proton bunches can be used to accelerate electrons over tens of meters with gradients on the order of one GeV/m.

The drive bunch energy (in Joules) determines the distance over which wakefields can be sustained. Electron bunches and laser pulses carry tens of Joules of energy and experimentally proved to be able to increase the energy of witness particles by several to tens of GeV. For example, it was demonstrated that wakefields driven by short electron bunches can accelerate electrons from 42 to \SI{84}{GeV} in \SI{0.85}{m} \cite{SLAC}. Wakefields driven by a short laser pulse were used to accelerate electrons up to \SI{4}{GeV} in \SI{9}{cm} of plasma \cite{BELLA}. However, to reach TeV energies, staging of multiple acceleration stages is required. Staging is an unsolved challenge, and may lead to a significant reduction of the effective gradient \cite{staging,lindstrom}.

Proton bunches produced at CERN carry energies of tens to hundreds of kilo-Joules. They thus have the potential to drive wakefields over long enough distances to accelerate electrons up to TeV energies \cite{Allen} in a single plasma stage.

The maximum accelerating gradient can be estimated by the cold plasma wavebreaking $E_{WB}$ field and depends on the plasma electron density $n_{pe}$:
\begin{equation}
\label{eq:EWB}
E_{WB} = \frac{m_e \omega_{pe} e}{c} \approx  100 \frac{eV}{m} \sqrt{n_{pe}/\textrm{cm}^{-3}},
\end{equation}
where $\omega_{pe}=\sqrt{n_{pe}e^2/(\epsilon_0 m_e)}$ is the angular plasma electron frequency, $m_e$ the electron mass, $e$ the electron charge, $\epsilon_0$ is the vacuum permittivity and $c$ is the speed of light. To reach GeV/m accelerating gradients, the plasma electron density needs to exceed $10^{14}$ cm$^{-3}$, as given by Equation \ref{eq:EWB}. 

To excite plasma wakefields effectively, the drive bunch length $\sigma_z$ must be shorter than the plasma electron wavelength $\lambda_{pe} = 2\pi c/\omega_{pe} \propto 1/\sqrt{n_{pe}}$, as given by the relation $k_{pe}\sigma_z \simeq \sqrt{2}$  \cite{lintheory}, where $k_{pe}=\omega_{pe}/c$. The radial bunch size $\sigma_r$ must be such that $k_{pe}\sigma_r\leq 1$ to avoid the development of the current filamentation instability\cite{CFI}. Consequently, to excite GV/m wakefield amplitudes, $\sigma_z$ must be shorter than $\sqrt{2}c/\omega_{pe}$ and $\sigma_r$ smaller than $c/\omega_{pe}$. For a plasma electron density of $10^{14}$cm$^{-3}$, $c/\omega_{pe}\approx$\SI{0.5}{mm}. While available high-energy proton bunches can be focused tightly ($\sigma_r\sim 100$ $\mu$m), their $\sigma_z \approx$ 6-\SI{15}{cm} is much too long for effective wakefield excitation at these densities.

When a bunch propagates in plasma, it drives longitudinal and transverse wakefields according to the bunch and plasma parameters. The particles at a certain location in the bunch are subjected to the wakefield that the particles ahead created. If the bunch is relativistic, and has thus a large forward momentum $p_{\parallel}$, the particles move close to the speed of light $v_{\parallel}\approx c$, and the longitudinal wakefields do not change their velocity significantly ($\Delta v_{\parallel}\ll c$). The transverse momentum of the bunches is typically much smaller ($p_{\perp} < 10^{-5} p_{\parallel}$), and the transverse wakefields can change significantly the particles transverse momentum and thus also their propagation angle.

\begin{figure}[!h]
\centering
\includegraphics[width=0.75\columnwidth]{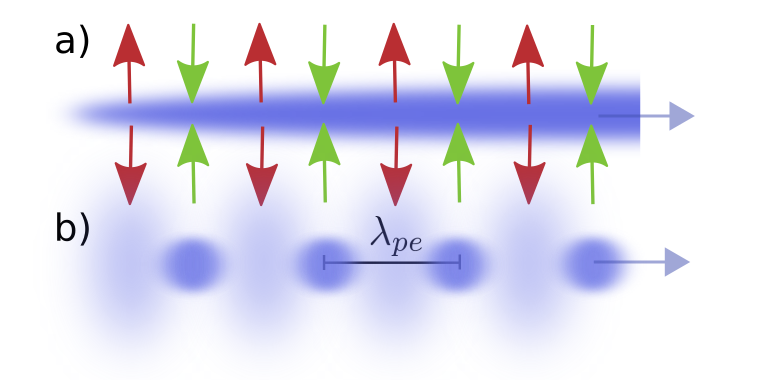}
\caption{a) Conceptual diagram of the transverse plasma wakefields that act on a proton bunch with a bunch length that is much longer than the plasma wavelength $\lambda_{pe}$. b) Formation of the on-axis density modulation (micro-bunches) as a result of the transverse plasma wakefields acting on the bunch.}
\label{fig:SSM}
\end{figure}

\begin{figure*}[!h]
\centering
\includegraphics*[width=\textwidth]{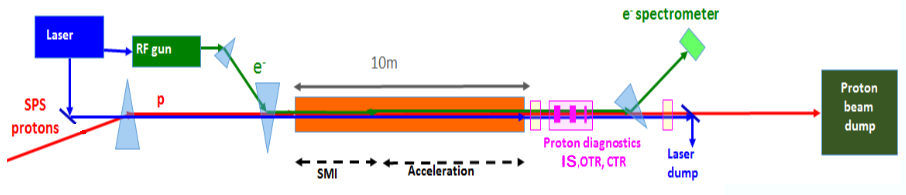}
\caption{Schematic layout of the main components of the AWAKE experiment. }
\label{fig:layout}
\end{figure*}

When the bunch is much longer than the plasma electron skin depth $c/\omega_{pe}$, the bunch particles experience either a focusing or defocusing force, depending on their location along the bunch (see Fig.\ref{fig:SSM}a). Where the bunch is defocused, its radial size increases and the proton density $n_p$ decreases. Where the bunch is focused, its radial size decreases and the proton density $n_p$ increases. Those regions then drive stronger wakefields ($\propto n_p$) creating a feedback loop for the self-modulation instability (SMI)\cite{SMI}. This process generates a periodic bunch density modulation that creates a train of micro-bunches, separated by the plasma wavelength $\lambda_{pe}$ (see Fig.\ref{fig:SSM}b). Each micro bunch fulfills the conditions for effective wakefield excitation ($k_{pe}\sigma_z \simeq \sqrt{2}$). Since they are spaced at $\lambda_{pe}$, they resonantly drive wakefields.

When the self-modulation process starts from noise, the wakefield phase and amplitude are not controlled. The undefined wakefield phase is a problem for externally injecting particles, because the proper phase for injection, accelerating and focusing appears randomly along the bunch. The solution is to seed the instability with a signal above noise level. The process is then called the seeded self-modulation (SSM) \cite{misha}. Seeding the instability fixes the phase and amplitudes of the wakefields so that they can be used to externally inject particles. Electrons are injected $\sim \sigma_z$ behind the seed point, where the wakefields reach a high amplitude.

In this article, we give a short update on the status of the AWAKE experiment. We start with a summary of the self-modulation experiments performed in 2017 and then discuss the plans and challenges for electron acceleration experiments foreseen for the near future.

\section{Status of the AWAKE Experiment}
The first run of the AWAKE experiment is separated into two phases:
\begin{itemize}
\item  Phase 1 (2016, 2017): Demonstration and understanding of the seeded self-modulation of a long ($\sigma_z \gg c/\omega_{pe}$) proton bunch in plasma.
\item Phase 2 (2018): Acceleration of electrons in plasma wakefields driven by a self-modulated proton bunch.
\end{itemize}

Figure \ref{fig:layout} shows the schematic layout of the experiment. A rubidium vapour source creates a uniform ($\delta n/n \leq \pm$\SI{0.25}{\%}) vapour density ($n$) with $1-10\times10^{14}$ atoms/cm$^3$ over a length of \SI{10}{m} \cite{erdem}. The system is open at both ends. Rubidium flows out of an aperture with a radius of \SI{5}{mm} and condenses in expansion volumes that are cooled below the rubidium freezing temperature (\SI{39}{°C}). The length of the density ramp at the beginning and end of the vapour source is on the order of the diameter of the opening apertures \cite{gennady}.

A \SI{120}{fs}, \SI{450}{mJ} laser pulse creates the plasma by ionizing the outermost electron of each rubidium atom over the total length of the vapour column and a radius of $\approx$ \SI{1}{mm}. 

The \SI{400}{GeV/c} proton bunch from the CERN Super Proton Synchrotron is overlapped in time and space with the laser pulse upstream the plasma. The rms proton transverse bunch size at the plasma entrance is $\sim 200\,\mu$ m, the rms length $\sigma_z =6-$\SI{15}{cm} and the bunch consists of $1-3\times10^{11}$ protons. The proton bunch and laser pulse traverse the vapor/plasma co-linearly. Placing the laser pulse inside the proton bunch creates a sudden onset of the plasma inside the proton bunch which acts as the seed for the SSM.

The goal of AWAKE run 1 is to show that the proton bunch self-modulated, that its parameters scale as expected and that the wakefields grow along the bunch and along the plasma. Therefore we measured downstream the plasma exit: 

\begin{itemize}
\item the on-axis proton bunch density as a function of time (with ps resolution): we image the optical transition radiation emitted by the proton bunch, when traversing a metallic foil, onto the slit of a streak camera \cite{karl}.
\item the radial displacement of the defocused protons: we measure the time-integrated transverse distribution of the proton bunch by imaging the light emitted by scintillating screens onto cameras\cite{marlene}.
\item the modulation frequency of the proton bunch: we detect and analyze the coherent transition radiation (CTR) emitted by the proton bunch when traversing a foil \cite{misha,falk}.
\end{itemize}

From the measurements, we show that the proton bunch self-modulated in the plasma and that the modulation is radial by observing periodic regions of low and high proton density on the streak camera images and observing protons that are defocused symmetrically around the bunch core on the time-integrated images.

We prove that the distance between the proton micro-bunches corresponds to the plasma electron wavelength $\lambda_{pe}$ by measuring the CTR frequency and by fourier transforming the streak camera images. We can additionally show that the proton micro-bunch position and thus the wakefield phase is constant with respect to the ionizing and seeding laser pulse. To determine the position of the ionizing laser pulse along the proton bunch on the streak camera images, we propagate a fraction of the laser light outside the plasma over the same distance as the ionizing laser pulse and send it onto the streak camera slit.

From the measurement of the radial displacement of the defocused protons we calculate the transverse momentum of the defocused protons. From their momentum change along the plasma we show that the plasma wakefield amplitude must have been higher than its initial seed level.

Preliminary results suggest that the proton bunch self-modulation developed successfully. Publications showing results and their analysis are being prepared and are expected to appear in the near future.

The seeded self-modulation experiments require a plasma with a uniform density, and a relativistic proton bunch to drive the wakefields. To demonstrate electron acceleration we additionally need to externally inject electrons into the focusing and accelerating phase of the wakefields.

The electrons are created by a photo-injector \cite{steffen}. We send a $\sim$\SI{100}{nJ}, $\sim$\SI{10}{ps} long UV-laser pulse onto a Caesium Tellurium (CsTe$_2$) photocathode, which produces the electrons. The bunch is then accelerated to first $\sim$\SI{5}{MeV} by a two and a half cells radiofrequency structure and then to 10-\SI{20}{MeV} in a one meter-long S-band booster before being transported to the plasma\cite{janet}. We aim to send a bunch with $10^9$ electrons, an rms bunch length $\sigma_z$ on the order of $\sim$\SI{1}{mm} and a rms bunch radius of $200\,\mu$m into the plasma.

\section{Prospects}

The electron gun and transport line have been installed in 2017 and are currently under commissioning. Electron acceleration experiments have not yet started but are foreseen for this year (2018). Once electrons, accelerated by plasma wakefields, exit the plasma we measure their energy with a magnetic spectrometer and a scintillating Lanex screen that is placed after the magnet. The magnet provides a magnetic field of 0.1-\SI{1.5}{T} over a length of \SI{1}{m} ($\int B dl = 0.1-1.5\,T\cdot m$). This enables us to measure accelerated electrons with energies ranging from \SI{30}{MeV} $-$ \SI{8.5}{GeV} on the spectrometer screen.

The main challenge of the acceleration experiment is to inject low-energy (\SI{16}{MeV}) electrons into the accelerating and focusing phase of the wakefields. At the beginning of the plasma, where the proton bunch is not self-modulated, and close to the seed point, the wakefields are focusing for particles with the same charge as that of the drive bunch (protons) and defocusing for the opposite charge (e.g. electrons). 

Simulation results \cite{AWAKE} suggest than injecting electrons co-linearly with the protons could cause them to be defocused: 
\begin{itemize}
\item first, there is a density ramp at the entrance of the vapor source\cite{gennady}; \item second, the proton bunch distribution evolves during self-modulation.
\end{itemize}
Both the change in plasma density and the bunch evolution changes the phase of the wakefields. Independent of their initial position along the wakefield, electrons therefore experience a defocusing field before significant acceleration can occur. Due to the electrons low energy ($\sim$\SI{10}{MeV}) the initial amplitude of the seed transverse fields of $\sim$\SI{10}{MV/m} can defocus electrons over a short distance, causing them to radially exit the wakefields.

To avoid the defocusing, we transport the electron bunch along a trajectory initially parallel to that of the proton bunch. We then use a corrector magnet to give their trajectory an angle, so that the electron bunch crosses that of the proton bunch at some distance into the plasma, where modulation of the bunch density has started and transverse wakefields are periodically focusing and defocusing.

For a first step, the goal is to inject enough electrons into the accelerating and focusing phase of the wakefield, so that we can detect them on the electron spectrometer (see Fig. \ref{fig:layout}). Since the proton bunch radius $\sigma_r \approx 200\,\mu$m and the radial extent of the wakefields $c/\omega_{pe} \sim 0.5-$\SI{0.15}{mm} is small we need to control and predict the electron bunch trajectory precisely. Since the electron bunch is long ($\sim \lambda_{pe}$), precise timing ($<$\SI{10}{ps}) of the electron bunch with respect to the wakefield phase is not necessary.

Due to the complexity of the vapor source, we are not able to have any bunch diagnostics inside and at the entrance and exit of the plasma. To predict the electron bunch position and radius at the proton-electron bunch crossing point, we measure the electron bunch position and transverse shape $\sim$\SI{1.5}{m},  $\sim$\SI{2}{m} and $\sim$\SI{4}{m} upstream of the vapor source entrance. 

\begin{figure}
\centering
\includegraphics[width=0.75\columnwidth]{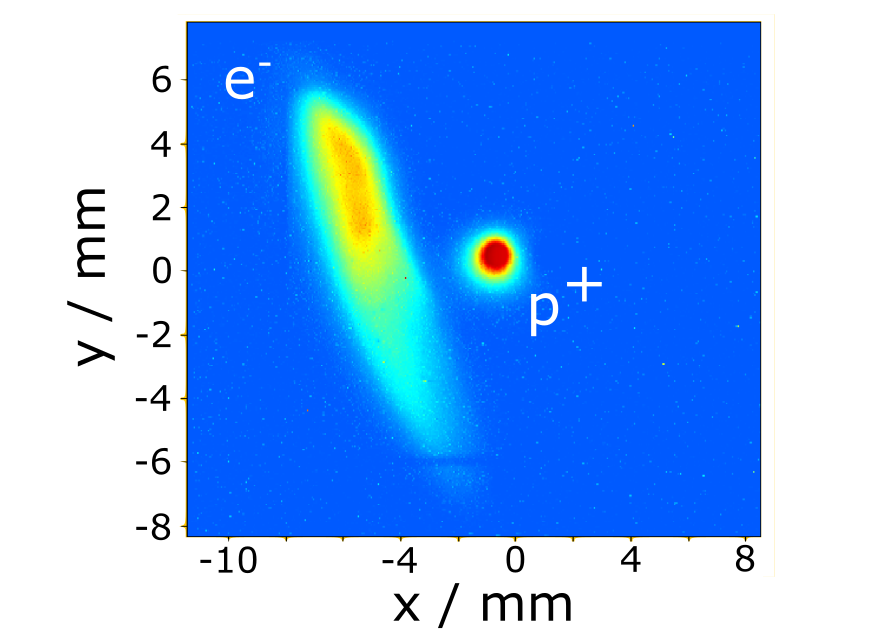}
\caption{Measurement of the transverse electron bunch distribution in presence of the proton bunch upstream the plasma entrance on a scintillating Chromox screen during the electron beam commissioning period in 2017. Note that the light emitted by the proton bunch locally saturates the camera pixels and that the electron bunch is not optimized.}
\label{fig:proelec}
\end{figure}
We insert either a \SI{1}{mm} thick scintillating Chromox (Al$_2$O$_3$:Cr$_2$0$_3$) screen or a \SI{0.2}{mm} thick silver coated silicon wafer into the beam path and image the luminescent or optical transition radiation onto cameras. This gives us the time-integrated transverse bunch profile at the screen location. Figure \ref{fig:proelec} shows such a measurement image at the screen $\sim$\SI{4}{m} from the plasma entrance. We can observe the electron and proton bunch at the same time, though the proton bunch appears saturated on the image. The electrons can only be measured at one screen position at a time, as at this low energy they acquire a large diverging angle due to scatterings in the first screen material.

In order to determine whether electron injection and trapping occurs, we measure the radiation produced by non trapped electrons when hitting the beam-pipe or vapor source walls. We are currently installing 12 electron beam-loss monitors around and upstream the vapor source. When electrons hit surrounding material, they scatter and produce X-rays. Part of the X-rays and scattered electrons can exit the system and deposit energy in the scintillating material inside the detectors. Preliminary FLUKA \cite{FLUKA} simulations suggest that we detect a signal if a minimum of $10^3$ electrons hit the wall over an area of $\sim 1\,\textrm{mm}^2$. Note that the scattered electrons as well as the X-rays spread over a large area.

\subsection{AWAKE run 2}

The goal of AWAKE run 1 is to accelerate low energy electrons to GeV energies. Since the electron bunch is longer than the wakefield period, only a fraction of the electrons will be trapped and thus the total amount of accelerated charge is small.

The next big step for AWAKE is to demonstrate that we can control the parameters of the accelerated electron bunch to the level where it can be used for first applications. For that, the electron bunch must have: a micron-level normalized emittance, a percent level relative energy spread and high charge. This is the goal of AWAKE run 2 that is scheduled from 2021 onwards. For high-energy physics applications we also need to demonstrate that we can produce a bunch with energies of tens of GeV or higher.

The idea for AWAKE run 2 is to inject electrons after the proton bunch self-modulation saturated and the wakefield phase stabilized, i.e. 4-\SI{10}{m} after the plasma entrance. For that, one can for example locally decrease the electron density or make a gap between the two plasmas. The distance between the two sections must be kept small to minimize the proton bunch evolution between the plasmas and must accommodate the electron injection.

Conceptually, this idea allows to operate with two plasma sections that could be of a different kind. The second source could be of a scalable technology, extendable to hundreds of meters. A plasma source that produces a uniform electron density on the order of $10^{14}$ cm$^{-3}$ over $\sim$\SI{100}{m} distance scales is currently developed.

\section{Summary and Conclusions}
In this article, we describe the AWAKE experiment and its goal of accelerating electrons in plasma wakefields driven by a relativistic, self-modulated proton bunch. We describe how we show that the bunch self-modulates successfully over the \SI{10}{m} of plasma: by the measurement of the on-axis proton density, the transverse momentum of the defocused protons and CTR emitted by the self-modulated bunch we show that the SSM develops, the modulation frequency corresponds to the plasma frequency and that the wakefields amplitude grow along the bunch and plasma. 

In 2018 AWAKE aims to demonstrate electron acceleration. The electron source and transport line have been installed and are currently under commissioning. The main challenge of the electron acceleration experiment is to inject electrons into the focusing and accelerating phase of the wakefield. We aim to inject electrons at an angle with respect to the proton bunch trajectory and we installed diagnostics around the entrance and along the plasma to investigate the injection process. 

The goal for run 2, which is scheduled from 2021 onwards, is to produce a high quality, controllable electron witness bunch.

\null

\end{document}